\documentclass[twocolumn]{aastex63}

\usepackage{graphics,epsf}
\usepackage{amsmath}                % American Mathematical Society package
\usepackage{amsfonts}               % American Mathematical Society fonts
\usepackage{amssymb}                % American Mathematical Society symbol
\usepackage{epsfig}                 % EPS figures
\usepackage{float}
\usepackage{color}
\usepackage{tabularx}
\usepackage{comment}
\usepackage{makecell}

\def \s{~\rm{s}}

\def \K{~\rm{K}}

\def \g{~\rm{g}}
\def \G{~\rm{G}}

\def \erg{~\rm{erg}}

\def \Myr{~\rm{Myr}}

\usepackage{xcolor}
\definecolor{redak}{rgb}{0.9,0.15,0.05}
\begin{document}
\title{The implications of Ultra-Faint Dwarf Galaxy Reticulum II on the common envelope jets supernova r-process scenario}
% OLD \title{Towards a paradigm change in the main heavy r-process nucleosynthesis sites}

\author{Aldana Grichener}
\affiliation{Department of Physics, Technion, Haifa, 3200003, Israel; aldanag@campus.technion.ac.il; soker@physics.technion.ac.il}

\author[0000-0003-0375-8987]{Noam Soker}
%\affil{Departmeמt of Physics, Technion, Haifa 3200003, Israel}
%%% \affiliation{Department of Physics, Technion, Haifa, 3200003, Israel; aldanag@campus.technion.ac.il; soker@physics.technion.ac.il}

%% \author{Rbert T. Fisher}
%% \affiliation{Department of Physics, University of Massachusetts  Dartmouth, 285 Old Westport Road, North Dartmouth, MA 02740, USA; rfisher1@umassd.edu}

\begin{abstract}
We show that the common envelope jets supernova (CEJSN) r-process scenario is compatible with very recent observationally determined properties of the stars in the ultra faint dwarf (UFD) galaxy Reticulum II that are strongly enhanced in r-process elements. These new results, like efficient mixing of the r-process elements in the Reticulum II galaxy, have some implications on the CEJSN r-process scenario for UFD galaxies. In particular, the energetic jets efficiently mix with the common envelope ejecta and then with the entire interstellar medium of Reticulum II. The compatibility that we find between the scenario and new observations suggests that the CEJSN r-process scenario supplies a non-negligible fraction of the r-process elements.
\end{abstract}


% ==========================================================
\section{Introduction}
\label{sec:Introduction}
% ==========================================================

There are several theoretically proposed r-process nucleosynthesis sites, including neutrino-driven wind in ordinary core-collapse supernovae (CCSNe; e.g., \citealt{ArconesThielemann2013}), jets in CCSNe (e.g., \citealt{Winteleretal2012}), neutron stars (NSs) merger (e.g., \citealt{Kasenetal2017}) including short-delay mergers (e.g., \citealt{BeniaminiPiran2019}), collapsars (e.g., \citealt{Siegeletal2019}), and common envelope jets supernovae (CEJSN; e.g., \citealt{GrichenerSoker2019a}). It is not clear yet which are the dominant r-process sites (e.g., \citealt{Kobayashietal2020, Kobayashi2022, VanderSwaelmenetal2022}). 

The properties of r-process enriched stars in ultra-faint dwarf (UFD) galaxies constrain and challenge these r-process nucleosynthesis scenarios. For example, the metal poor stars of Reticulum II that are highly r-process enriched (\citealt{Jietal2016, Roedereretal2016}) imply short delay from star formation to the r-process event(s). Moreover, it is possible that the low-metalicity r-process stars in the Milky Way Galaxy originated from UFD galaxies that the Galaxy accreted (e.g., \citealt{Hiraietal2022}). 
 
% ==========================================================
\section{The r-process in the Reticulum II galaxy}
\label{sec:Reticulum}
% ==========================================================

In very recent studies \cite{Jietal2022} and \cite{Simonetal2022} explore the distribution of r-process nucleosynthesis in the UFD galaxy Reticulum II. We list some of their findings in the first column of Table \ref{table:Reticulum}, and the implications of these findings to a potential r-process site in the second column. We point out that the CEJSN r-process scenario, in which a cold NS accretes mass from the core of a giant star and launches neutron-rich jets, can account for, or at least is compatible with, the properties of r-process nucleosynthesis in Reticulum II, as we list in the third column of Table \ref{table:Reticulum}. 
% TTTTTTTTTTTTTTTTTTTTTTTTTTTTTTTTTTTTTTTTTTTTTTTTTTTTTT
\begin{table*}
\begin{center}
\caption{Findings of \cite{Jietal2022} and \cite{Simonetal2022} regarding r-process distribution in Reticulum II and their implications on the CEJSN r-process scenario.}
\setlength\tabcolsep{3 pt}
\begin{tabular}{|| c c c c ||}
 \hline
\makecell{Property} & \makecell{Implication} & \makecell{Explanation by the \\ CEJSN 
r-process scenario } & \makecell{New implications \\ to the
CEJSN \\ r-process scenario
 } \\ 
  & & & \\ %[0.5ex] \\ 
 \hline
 \makecell{ $\approx 28\%$ of stars have \\ no r-process \\ and formed \\ within $500 \Myr $ \\ $ \pm 200 \Myr
^{\textcolor{blue}{[Si22]}}$. } 
& \makecell{1. Substantial star  \\ formation before \\ the r-process event, \\ that must take \\ place no later \\ than 
$500 \Myr$ \\ $\pm 200 \Myr
^{\rm \textcolor{blue}{ [Si22]}}$.}
& \makecell{Delay time from \\ binary formation \\ to CEJSN r-process \\ enrichment $\simeq 10-$ \\ $30\Myr^{\textcolor{blue}{[GS19b]}}$,\\ leaving time for \\ star formation. 
} 
& \makecell{The binary \\ progenitor of \\ the r-process \\ event  was \\ formed at the \\ end of the first \\ star formation \\ episode.  
 } \\
 \hline
 \makecell{A small variation \\ in $\rm [Ba/H]$ 
among \\ r-process stars$^{\rm \textcolor{blue}{ [Ji22]}}$.  } 
& \makecell{2. Almost \\  uniform mixing \\ of r-process in \\ the galaxy. }
& \makecell{Wide jets (disk- \\
wind)$^{\rm \textcolor{blue}{[S22]}}$ that \\ efficiently mix \\ with the ejected \\ common envelope.
  } 
& \makecell{ The very \\ energetic explosion \\ induces turbulence \\ in the entire \\ ISM of the \\ UFD galaxy that \\ mixes the ejecta \\ with the ISM.  
 } \\
 \hline
 \makecell{A small variation \\ in $\rm [Ba/H]$ 
among \\ r-process stars$^{\rm \textcolor{blue}{[Ji22]}}$.  } 
& \makecell{3. No star \\ formation from \\ r-process event \\ to next star \\ formation episode. }
& \makecell{ The very energetic \\ CEJSN
explosion, \\ $\ga 10^{51} \erg ^{\rm \textcolor{blue}{ [So19]}}$ \\ excites shock waves \\ in the ISM that \\  terminate star \\ formation, allowing \\ time for mixing.  
 }
& \makecell{} \\
 \hline
 \makecell{Large variations \\ in [Fe/H]. } 
& \makecell{4. CCSNe and/or \\ SNe Ia after \\ the mixing of \\ r-process and \\ before or during \\ the next star  \\ forming episode.
 }
& \makecell{ } 
& \makecell{CCSNe and even  \\ SNe Ia occur \\ during the post-\\ mixing star \\ formation episode. 
 } \\
 \hline
\makecell{High r-process \\ enrichment. } 
& \makecell{5.The r-process \\ event takes place \\ inside the galaxy.\\ Namely, short delay \\ distance
 $D_{\rm nk}$.
}
& \makecell{$D_{\rm nk} \lesssim R_{\rm DG}$ \\ leaves most \\ r-process isotopes \\ inside this UFD \\ galaxy$^{\rm \textcolor{blue}{ [GS19b]}}$.
 } 
& \makecell{} \\
 \hline
 \makecell{Total number of \\ CCSNe in \\ Reticulum II  \\ $\approx 180^{\rm \textcolor{blue}{[Si22]}}$. About \\ third in the \\ first episode \\ of non-enriched \\ stars.
 } 
& \makecell{6. One r-process \\ event per $\approx 100$ \\ CCSNe in very  \\ low-metallicity \\ stars.
 }
& \makecell{The expected NS-\\core merger to \\ CCSN ratio is \\ $~1\%^{\rm \textcolor{blue}{ [Gr23]}}$ 
 } 
& \makecell{ } \\
 \hline
\end{tabular}
\end{center}
%\centering
{\textbf{Acronyms:} CEJSN: common envelope jets supernova; ISM: interstellar medium; UFD: ultra-faint dwarf; CCSN: core collapse supernova; SN Ia: type Ia supernova; NS: neutron star.
\\ \textbf{Definitions:} $D_{\rm nk}$: distance from star formation zone to r-process event; $R_{\rm DG}$: the radius of the UFD galaxy.
\\ \textbf{References:} GS19b: \cite{GrichenerSoker2019b}; Ji22: \cite{Jietal2022}; Gr23: Grichener 2023, prepeint; 
Si22: \cite{Simonetal2022};  So19:\cite{Sokeretal2019}; S22: \cite{Soker2022}. 
 
}
\label{table:Reticulum}
%\end{center}
\end{table*}
% TTTTTTTTTTTTTTTTTTTTTTTTTTTTTTTTTTTTTTTTTTTTTTTTTTTTTT

Note that in addition to the delay time from star formation to the r-process event, we follow \cite{GrichenerSoker2019b} and define the delay distance $D_{\rm nk}= v_{\rm nk} \left( t_{\rm 0-rp} - t_{\rm kick} \right)$,  where $t_{\rm 0-rp}$ is the delay time from star formation to r-process nucleosynthesis, and $t_{\rm kick} \simeq 10-30 \Myr$ is the time from star formation to the time the system suffers the natal kick. This is the distance from the formation place of the binary system to the r-process event.

The very recent studies by \cite{Jietal2022} and \cite{Simonetal2022} have some new implications to the CEJSN r-process scenario with respect to the UFD galaxy Reticulum II, as we present in the fourth column of Table \ref{table:Reticulum}. These implications are not applicable to the r-process in the Milky Way Galaxy, that the CEJSN r-process scenario can account for \citep{GrichenerSoker2019b, Gricheneretal2022}.   

% ==========================================================
\section{Conclusions}
\label{sec:Conclusions}
% ==========================================================

Our conclusion is that the CEJSN r-process scenario might account for the new findings regarding the r-process properties of the UFD galaxy Reticulum II. In addition to the already mentioned characteristics of this scenario that account for the general r-process properties (e.g., \citealt{GrichenerSoker2019a}), there are new implications that are applicable to the Reticulum II and possibly other UFD galaxies, but not for the Milky Way Galaxy. 

First, a very efficient mixing of the r-process isotopes must take place in the entire UFD galaxy (implication 2 in Table \ref{table:Reticulum}). This is not the case with the Milky Way. We attribute this to the energetic jets of the CEJSN event that first mix the jets with the CEJSN ejecta, and later with the entire interstellar medium (ISM). The explosion energy cannot efficiently mix the ejecta in a large galaxy as the Milky Way. 

The required rate of about one event per 100 CCSNe (implication 6 in Table \ref{table:Reticulum}) is compatible with new results of the CEJSN rate (Grichener 2023, preprint). 

Despite the compatibility of the CEJSN r-process scenario with the properties in Reticulum II and other advantages of this scenario, we do not claim that this is the only r-process site, but rather that it contributes significantly to r-process nucleosynthesis.

\end{document}